\documentclass[3p,twocolumn,sort&compress, fleqn]{elsarticle}
\usepackage{amsmath, bm}
\usepackage{amssymb}
\usepackage{amsmath}
\usepackage{braket}
\usepackage{listings}
\usepackage{cprotect}
\usepackage{lmodern}
\usepackage{ulem}

\journal{Computer Physics Communications}

\newcommand{\ham}{\mathcal{H}}

\usepackage{natbib}
\usepackage{hyperref}

\hypersetup{
        colorlinks=true,
}
\usepackage{listings}

\long\def\beginmypgfpdfnamed#1#2\endmypgfpdfnamed{\includegraphics{#1}}

\graphicspath{{./pyfig/}}

\usepackage{color}

\begin{document}

\begin{frontmatter}

\title{DSQSS: Discrete Space Quantum Systems Solver}

\author[a]{Yuichi Motoyama}\ead{y-motoyama@issp.u-tokyo.ac.jp}
\author[a]{Kazuyoshi Yoshimi}
\author[b,c]{Akiko Masaki-Kato}
\author[a]{Takeo Kato}
\author[a]{Naoki Kawashima}

\address[a]{Institute for Solid State Physics, University of Tokyo, Chiba 277-8581, Japan}
\address[b]{Computational Condensed Matter Physics Laboratory, RIKEN
Cluster for Pioneering Research (CPR), Wako, Saitama 351-0198, Japan}
\address[c]{Computational Materials Science Research Team, RIKEN
Center for Computational Science (R-CCS),  Kobe, Hyogo 650-0047,
Japan}

\begin{abstract}
The Discrete Space Quantum Systems Solver (DSQSS) is a program package for solving quantum many-body problems defined on lattices. The DSQSS is based on the quantum Monte Carlo method in Feynman's path integral representation and covers a broad range of problems using flexible input files that define arbitrary unit cells in arbitrary dimensions and arbitrary matrix elements representing the interactions among an arbitrary number of degrees of freedom. Finite temperature calculations of quantum spin and the Bose-Hubbard models can be performed by specifying parameters such as the number of dimensions,  the lattice size, coupling constants, and temperature. The present paper details the use of DSQSS and presents a number of applications thereof.
\end{abstract}

\begin{keyword}
Quantum Monte Carlo method, Quantum spin model, Bose-Hubbard model
\end{keyword}

\end{frontmatter}
{\bf PROGRAM SUMMARY}

\begin{small}
\noindent
{\em Program Title:} DSQSS \\
{\em Journal Reference:}                                      \\
{\em Catalogue identifier:}                                   \\
{\em Licensing provisions:} GNU General Public License version 3\\
{\em Programming language:} \verb*#C++# and  \verb*#python#\\
{\em Computer:} PC, cluster machine\\ 
{\em Operating system:} UNIX like system, tested on Linux and macOS\\ 
{\em Keywords:} Quantum Monte Carlo method, Quantum spin model, Bose-Hubbard model.\\ 
{\em Classification:} 6.5 Software including Parallel Algorithms. 7.7 Other Condensed Matter inc. Simulation of Liquids and Solids. \\ 
{\em External routines/libraries:}  MPI, NumPy, SciPy \\
{\em Nature of problem:} Finite temperature calculation of quantum spin model and Bose-Hubbard model.\\
{\em Solution method:} Path-integral Monte Carlo method with the directed-loop algorithm. \\
\end{small}

\section{Introduction}
The Quantum Monte Carlo (QMC) method based on the Feynman path integral is a powerful and widely used tool for studying strongly correlated quantum systems ~\cite{Suzuki1976,Gubernatis2016,StronglyCorrelatedSystems}.
For lattice models, the QMC algorithm implemented with a nonlocal loop update~\cite{Evertz1993,Prokofev1998,Sandvik1999,Todo2001,Evertz2003,Kawashima2004} works very efficiently as long as no sign problem appears.
For quantum spin systems without frustration and for bosonic systems, both of which are free from the sign problem, a QMC simulation can be performed for a lattice with a large number of sites and can clarify the detailed phase diagram of the system and the quantum critical behavior near a phase transition.
QMC simulations have successfully been applied to quantum Heisenberg models~\cite{Yasuda2005}, the SU($N$) Heisenberg model~\cite{KawashimaTanabe_2007, Harada2013, Motoyama_2018}, the hard-core boson model~\cite{Wessel2005}, and the Bose-Hubbard model~\cite{Kato2009,Trotzky2010,Ohgoe2012,Kato2014}.

In the present paper, we describe the open-source Discrete Space Quantum Systems Solver (DSQSS) version 2 code, which implements a continuous-time path-integral QMC (PIQMC) algorithm based on a directed-loop algorithm~\cite{Syljuasen2002,Syljuasen2003}.
This code is composed of two subpackages: the Discrete Space Quantum Systems Solver/directed-loop algorithm (DSQSS/DLA) and  the Discrete Space Quantum Systems Solver/parallelized multiple-worm algorithm (DSQSS/PMWA).
The DSQSS/DLA implements an on-the-fly directed-loop update algorithm~\cite{Pollet2007,Kato2007,Kato2009} and can perform QMC simulation for quantum spin/boson models on an arbitrary lattice.
The DSQSS/PMWA is a unique code that implements a parallelizable multi-worm QMC algorithm, which can be applied to extremely large systems based on the spin-$1/2$ XXZ model or the hard-core boson model~\cite{Masaki-Kato2014}.

The DSQSS has several new features in comparison with the preceding software, such as the ALPS code~\cite{Albuquerque2007,Bauer2011,ALPS}. 
The present code can be installed in a simple manner and requires only a user-friendly input file.
The DSQSS can also output imaginary-time correlation functions, which can be used to calculate dynamic structure factors by numerical analytic continuation.
The on-the-fly algorithm used in the DSQSS/DLA has an advantage in simulating dilute boson systems, which are usually encountered when discretizing continuum bosonic systems. 
The DSQSS/DLA also supports a wider range of models, including multi-body interactions beyond two-body interactions  and interactions that have a complicated representation in terms of SU(2) operators.

The nontrivial massive parallelization implemented in the DSQSS/PMWA provides a pathway for extremely large-size QMC simulations on massively parallel machines.

The remainder of this paper is organized as follows. 
In Section 2, we briefly explain the algorithm used in the present code. 
In Section 3, we describe the basic usage of the code. 
In Section 4, we give a few instructive examples to show the practical usage of the code.
We attempt only to present an outline of the usage of the package in the present paper. For fuller details, we refer the reader to the online manual~\cite{dsqss}.
Section 5 contains a summary of this paper.

\section{Algorithm} 

In this section, we briefly describe the algorithm used in the DSQSS.
We explain the three steps in the algorithm: How to construct a Monte Carlo (MC) configuration with a c-number weight, how to update the configuration, and how to evaluate physical quantities.
For more details, such as derivations of equations, readers can refer to relevant review articles~\cite{Kawashima2004} and/or textbooks~\cite{StronglyCorrelatedSystems, Gubernatis2016}.

\subsection{Path-integral representation}

In order to construct a MC configuration, we expand the partition function for the system described by the Hamiltonian $\ham$ at the inverse temperature $\beta$, $Z = \mathrm{Tr} \exp(-\beta \ham)$, as a summation of \textit{c}-numbers, $Z = \sum_c W(c)$, where $W(c)$ is the weight of configuration $c$.
In the DSQSS, $W(c)$ is constructed by the continuous time path-integral expansion with the interaction representation.
First, we split the Hamiltonian $\ham$ into the diagonal part $\ham_0$ and the off-diagonal part $V = \sum_p V_p$, where $V_p$ is a local operator with $p$ specifying a local set of sites (e.g., a nearest neighbor pair, $p=\langle ij \rangle$).
For the Heisenberg model, we choose a basis set, where the $z$ components of the local spin operators $S^z$ on all sites are simultaneously diagonalized, $\ket{\phi} = \otimes_i \ket{\sigma_i}$, where $\ket{\sigma_i}$ is an eigenstate satisfying $S^z_i\ket{\sigma_i} = \sigma_i\ket{\sigma_i}$ at the $i$-th site.
In terms of this basis set, the diagonal part of the Hamiltonian consists of the Ising terms with the Zeeman term $\ham_0 = J_z \sum_{\langle ij \rangle} S^z_i S^z_j - h\sum_i S_i^z$ and the off-diagonal part is the XY terms $V = \sum_{\langle ij\rangle} V_{ij} =  \left(J_{xy}/2\right)\sum_{\langle ij \rangle}(S^+_i S^-_j + S^-_i S^+_j)$.
By using $\ham_0, V_p$ and $\ket{\phi}$,
the partition function $Z$ can be expanded as follows:
\begin{equation}
\begin{split}
    Z &= \sum_{n=0}^\infty \sum_{\{\phi_i\}}\sum_{\{p_k\}} \int_0^\beta d\tau_n \int_0^{\tau_n} d\tau_{n-1} \cdots \int_0^{\tau_2} d\tau_1 \\
    & \quad \times \prod_{k=1}^{n}
    e^{-(\tau_{k+1}-\tau_k)E^0_{k+1}}
    \Braket{\phi_{k+1}|(-V_{p_k})|\phi_k}
\end{split}
\label{eq:expansion}
\end{equation}
where $E^0_k$ is the eigenenergy of $\ket{\phi_k}$ with respect to the diagonal part $\ham_0$, and periodic boundary conditions along the imaginary time axis, $\ket{\phi_{n+1}} = \ket{\phi_1}$ and $\tau_{n+1} = \beta + \tau_1$, hold due to the trace operation.
In this representation, a MC configuration $c$ is specified by a set of variables: the number of operators $n$, the series of states $\{\phi_k\}$, the series of imaginary times $\{\tau_k\}$, and the series of indices $\{p_k\}$.
The weight of this MC configuration is given by
\begin{equation}
\begin{split}
    & W(n,
    \{\phi_k\},
    \{\tau_k\}, \{p_k\}) 
    \\
    & =
    \prod_{k=1}^n 
    e^{-(\tau_{k+1}-\tau_k)E^0_{k+1}}
    \Braket{\phi_{k+1}|(-V_{p_k})|\phi_k}.
\end{split}
\label{eq:weight}
\end{equation}

\begin{figure}
    \centering
    \includegraphics[width=\linewidth]{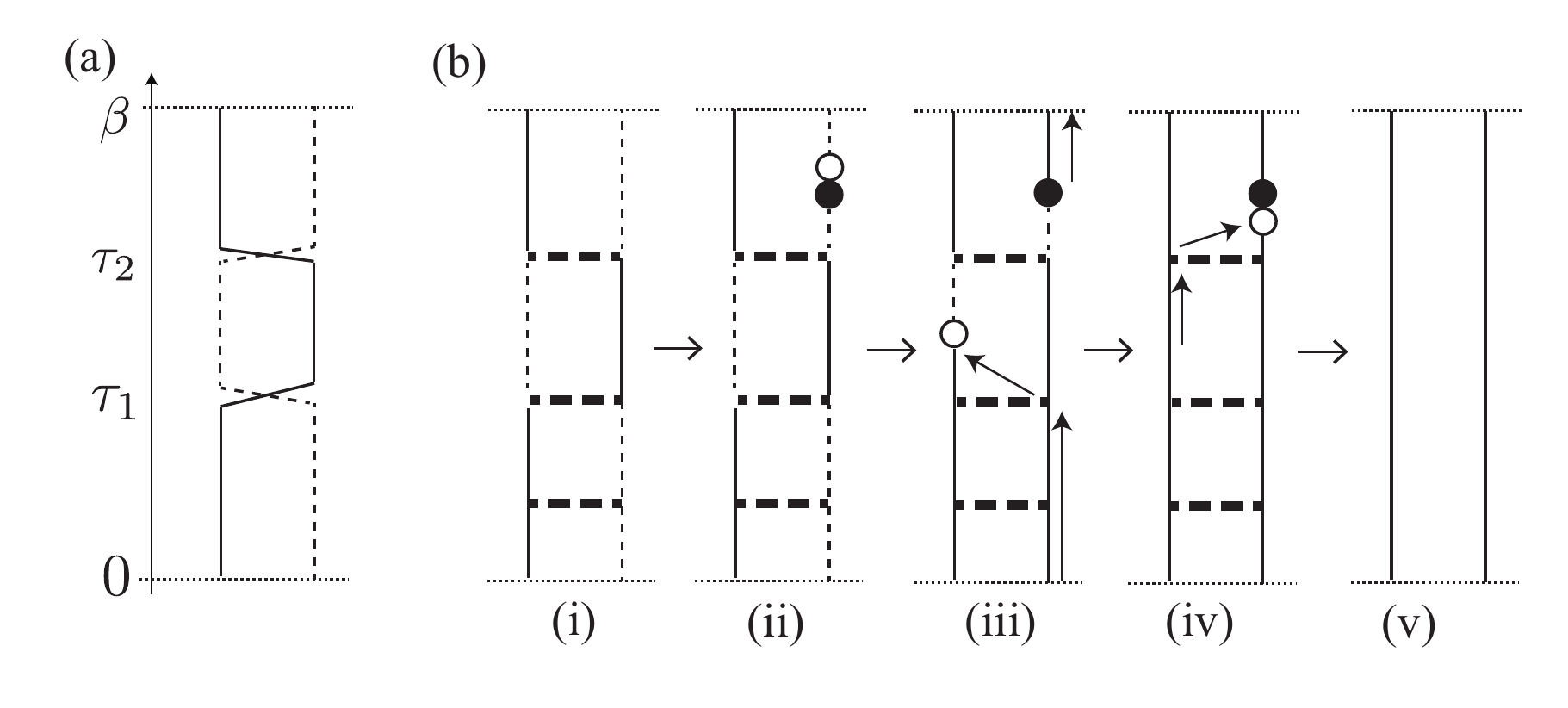}
    \caption{
    (a) Example of a world-line Monte Carlo (MC) configuration for a spin dimer. The solid and dashed lines represent the up-spin and down-spin states, respectively.
    (b) Example of the update process for the MC configuration (a):
    (i) Vertex insertion, (ii) creation of a worm head-tail pair, (iii) movement of the worm head, (iv) return of the worm head to the position of the tail, and (v) removal of the vertices.
    The filled and empty circles indicate the off-diagonal operators, $S^+$ and $S^-$, respectively.
}
    \label{fig:world-line}
\end{figure}

As a simple example, let us consider a $S=1/2$ antiferromagnetic Heisenberg dimer described by $\ham = \vec{S}_1\cdot \vec{S}_2$.
An example of the MC configuration for this model is depicted in Fig. \ref{fig:world-line}(a).
The solid and dashed lines represent the up-spin states $\ket{\uparrow}$ and down-spin states $\ket{\downarrow}$, respectively.
In Fig. \ref{fig:world-line}(a), the initial state at $\tau=0$ is $\ket{\phi_1} = \ket{\uparrow}_1\ket{\downarrow}_2$, and the imaginary-time evolution from $\tau = 0$ to $\tau = \beta$ under the Hamiltonian $\ham_0 = S^z_1 S^z_2$ is perturbed by $S^-_1 S^+_2$ at $\tau_1$ and by $S^+_1 S^-_2$ at $\tau_2$.
Thus, the MC configuration can be represented as lines, which are trajectories of ``particles'' in the $d+1$-dimensional Euclid spacetime.
This is why the configuration is referred to as a ``world-line'' and the PIQMC is also referred to as a world-line QMC.

\subsection{Update algorithm}

In MC sampling, the world-line configuration is updated based on the balance condition.
In order to realize efficient updates, several sophisticated update algorithms, such as the loop algorithm, the worm algorithm, and the directed-loop algorithm (DLA), have been developed.
In the present program package, the DSQSS/DLA implements the directed-loop algorithm, whereas the DSQSS/PMWA implements the parallelized multiple worm algorithm (PMWA).
We will briefly review these two algorithms in subsequent subsections.

\subsubsection{Directed-loop algorithm }

The DLA is one of the variations of the worm algorithm, which is combined with the loop algorithm\footnote{
The DLA is regarded as a PIQMC version of the Wolff algorithm, while the loop algorithm is a PIQMC version of the Swendseng-Wang algorithm.}.
By adding a source term $\eta \sum_i Q_i$ to the Hamiltonian, the DLA introduces a pair of ``worm heads'', where $Q_i$ is an onsite off-diagonal operator, such as $S_i^x$ for quantum spin models.
Generally, the value of the conjugate field $\eta$ is arbitrary, and the MC configuration $c$ can include an arbitrary number of pairs of worm heads, $n_{w}$.
In the DLA, $\eta$ is chosen so that the probability of pair-annihilation is unity.
The probability of pair-creation is then determined from the detailed balance condition.
By this choice of $\eta$, the DLA constructs a ``worm update'', which is composed of creation, movement, and annihilation of a pair of worm heads.
This worm update changes world-line MC configurations efficiently.
Note that the MC configurations before and after the worm updates include no worm heads, whereas one pair of worm heads appears during the worm update.
An example of the worm update process for a spin dimer is shown in Fig. \ref{fig:world-line}(b).

\begin{figure}
    \centering
    \includegraphics[width=\linewidth]{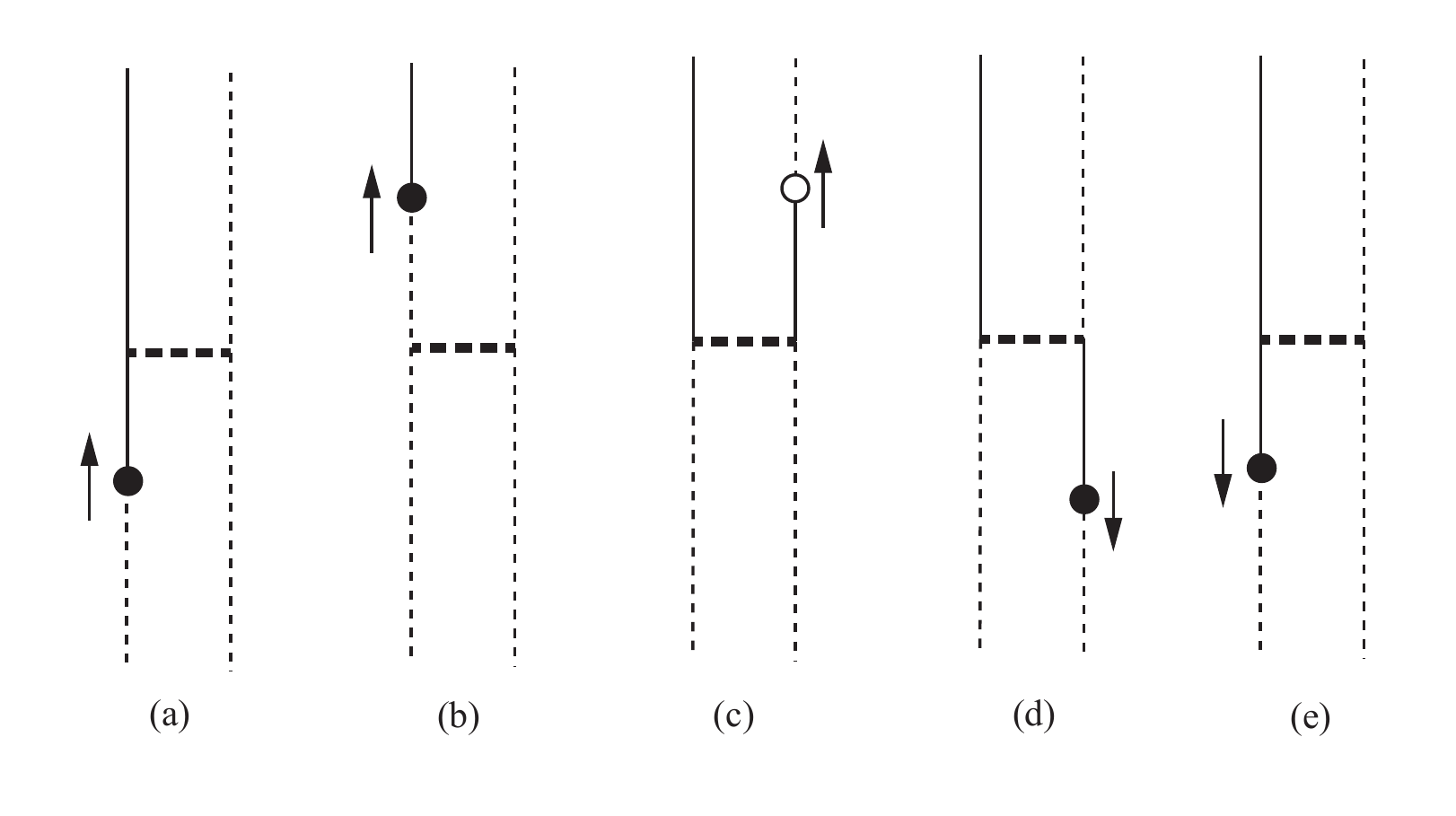}
    \caption{A worm head is scattered by a vertex. The arrow indicates the direction of the head. (a) Initial state. (b-e) Candidates for final state.}
    \label{fig:scattering}
\end{figure}

The details of the DLA are as follows~\footnote{The DSQSS/DLA requires the parameters described in this subsection, e.g., the density of the vertices and the transition probabilities for scattering, as inputs.
The DSQSS provides tools  for calculating these parameters from the local Hamiltonians and preparing the input files.}.
In the DLA, we first distribute ``vertices'' with a density according to the diagonal elements of the local Hamiltonians on a world-line configuration as in the loop algorithm.
Next, we choose a random spacetime point and try to create a pair of worm heads at this point.
The worm head has its traveling direction along the imaginary time axis and is straight until encountering a vertex.
When the worm head reaches a vertex, it is scattered, and travels along one of the ``legs'' of the vertex.
Figure ~\ref{fig:scattering} shows a two-site vertex case. Panel (a) is the initial state, and panels (b) through (e) are candidates for the final states after scattering.
The type of worm head (the operator that the head represents) can also be changed through scattering.
Note that for a binary state, such as $S=1/2$ spin systems, the worm head type is automatically determined from which leg the head is on (see Fig.~\ref{fig:scattering}).
The final states are chosen based on the scattering probabilities, which are determined from the weight of each state according to the balance condition.
The DSQSS/DLA implements the following methods for calculating scattering probabilities: the Metropolis-Hastings algorithm~\cite{Metropolis_1953, Hastings_1970}, the heat-bath algorithm~\cite{Creutz_1980}, the Suwa-Todo algorithm~\cite{Suwa2010}, and the reversible Suwa-Todo algorithm~\cite{Suwa2013, Todo2013}.
Finally, when a head returns to the other head, the pair is removed.\footnote{The DSQSS/DLA refers to the process from worm creation to worm annihilation as one MC cycle. 
Failure of worm creation is also referred to as one MC cycle. 
One MC sweep consists of $N_\text{cycle}$ MC cycles, where $N_\text{cycle}$ is determined in the initial MC cycles such that the expectation value for the movement distance for worm heads in one MC sweep is the volume of the spacetime, $N_\text{sites}\beta$.}.

If the diagonal interactions and/or the external field are very strong, a large number of vertices are present in the first step of the worm update, but most are not used because the worm head moves in a limited region of spacetime.
In order to avoid this inefficient procedure, the DSQSS/DLA implements an on-the-fly algorithm~\cite{Kato2007,Kato2009}, where vertices are inserted only when a non-trivial scattering event takes place.
This reduces computational and memory costs drastically and thus enables the DSQSS/DLA to deal with strongly correlated or diluted systems.

\subsubsection{Parallelized multiple-worm algorithm}

Since large-scale simulations on modern supercomputers use a large number of CPU cores, non-trivial parallelization of the algorithms is required.
One type of non-trivial parallelization is domain parallelization, where the whole spacetime is split into pieces, and a processor is assigned to each of the pieces~\cite{Todo_2019, ALPS_looper}.
However, domain parallelization of the DLA is not straightforward because only one worm head updates the state \textit{locally}.
In order to overcome this difficulty, a parallelized multi-worm algorithm (PMWA), which is an extension of the DLA for domain parallelization, has been proposed \cite{Masaki-Kato2014}.
The PMWA introduces a number of pairs of worm heads such that every domain contains a sufficient number of worm heads.
Each CPU core creates, annihilates, and moves worm heads in the corresponding domain independently, and sometimes communicates with the cores controlling the neighboring domains for exchanging worm heads across the boundaries.
The domain parallelization used in the PMWA shows good scaling with respect to the number of CPU cores \cite{Masaki-Kato2014}.

One MC cycle in the PMWA is composed of a vertex phase and a worm phase.
In the vertex phase, vertices are distributed in the same way as in the DLA. (The DSQSS/PMWA implements a not-on-the-fly version.)
The worm phase processes many worm heads in three steps: (i) creation/annihilation, (ii) worm scattering, and (iii) domain-boundary update.
For efficient parallelization, each domain must have an approximately equal number of worm heads. In order to ensure this, the source field $\eta$ must be sufficiently large \footnote{
Note that for the XXZ model with antiferromagnetic XY-like interactions, a uniform transverse field causes the negative sign problem, even if the lattice is bipartite. 
A staggered field, on the other hand, does not cause this problem, and thus the DSQSS/PMWA deals with such a model.}.
Therefore, it is practically impossible to update the world-line configurations using only the MC configurations with no worm heads or one pair of worm heads, as in the DLA.
The user needs to measure physical quantities with finite values of $\eta$ and to extrapolate them to $\eta=0$.

\subsection{Measurements of quantities}
In order to evaluate the expectation value of an observable $A$ from a series of configurations $\{c_i\}$, we should define a suitable estimator $A(c)$ and evaluate $\braket{A} = \sum_i A(c_i) / N$.
In the DSQSS, four kinds of physical quantities can be calculated: (i) observables for diagonal operators, (ii) energy and specific heat, (iii) helicity modulus, and (iv) observables for off-diagonal operators. For details of the method for calculating these physical quantities, see Appendix A.

\section{Basic usage}
\subsection{Installation}
Installation of the DSQSS can be performed using the following procedure.
In the following, we assume that the user is in the root directory of the DSQSS.

\begin{verbatim}
$ mkdir dsqss.build
$ cd dsqss.build
$ cmake ../
$ make
\end{verbatim}
The default installation directory is \verb|/usr/local/bin|.
In order to change the installation directory,
add the cmake option \verb|-DCMAKE_INSTALL_PREFIX=/path/to/install/to|,
where \verb|/path/to/install/to| is the path for the installation directory.
Each binary file for the DSQSS will be made in the \verb|src| and \verb|tool| directories. 
In order to check whether the binary files are correctly created, type the following command:
\begin{verbatim}
$ make test
\end{verbatim}
After all of the tests have been passed, type the following command to install the files:
\begin{verbatim}
$ make install
\end{verbatim}

This command installs executable files into the \verb|bin| directory,
sample files into the \verb|share/dsqss/VERSION/samples| directory,
and the python package \verb|dsqss| into the \verb|lib| directory
under the installation path that was set previously.

One configuration file for setting environment variables to perform the DSQSS commands will also be created, and named \verb|share/dsqss/dsqssvar-VERSION.sh|.
Before invoking the DSQSS commands, load this file using the \verb|source| command as

\begin{verbatim}
$ source share/dsqss/dsqssvar-VERSION.sh
\end{verbatim}
\subsection{Usage of the DSQSS/DLA}
\begin{figure*}[tb!]
  \begin{center}
    \includegraphics[width=16.0cm]{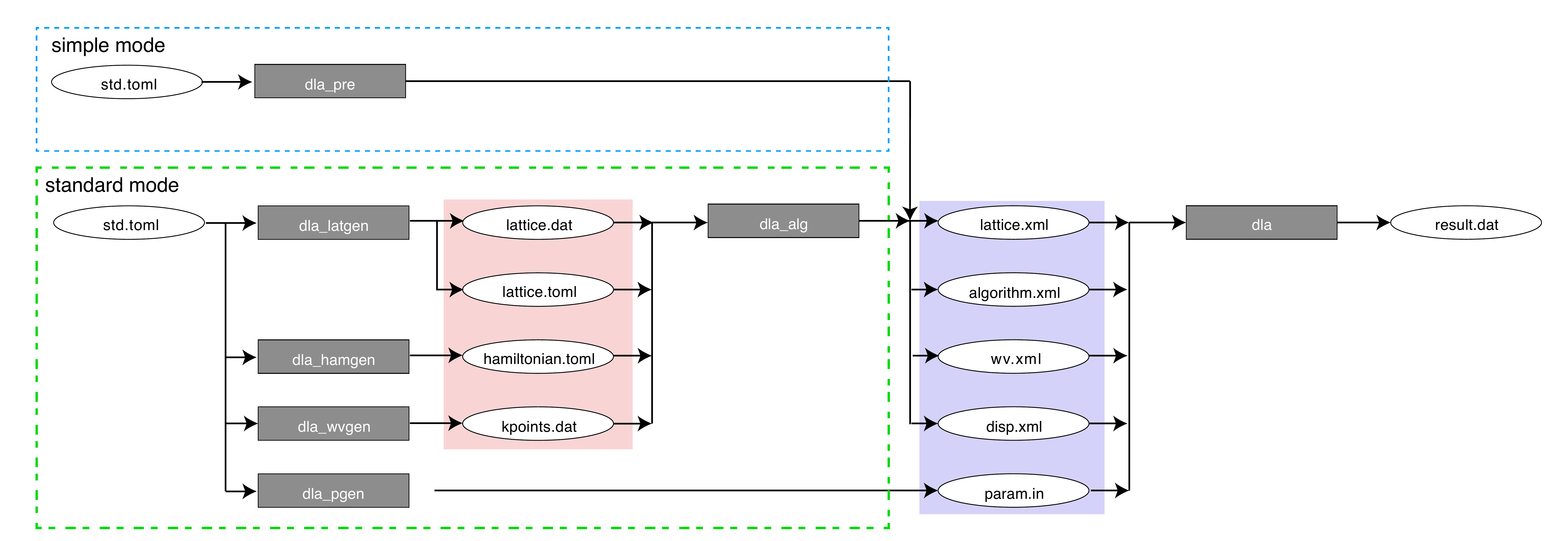}
    \cprotect\caption{Schematic flow of simple and standard modes of the Discrete Space Quantum Systems Solver/directed-loop algorithm (DSQSS/DLA). Ellipses are files and rectangles are tools.}
    \label{fig:flow_dsqss}
  \end{center}
\end{figure*}

The DSQSS/DLA calculations involve three steps:
(i) preparing input files, (ii) performing QMC calculations, and (iii) analyzing results.
For making the DSQSS/DLA input files, there are two modes: a simple mode and a standard mode, as shown in Fig. \ref{fig:flow_dsqss}.
The simple mode offers the easiest workflow. In this mode, users can simulate a predefined model on a predefined lattice from one text file. 
In the standard mode, users can define their own models and lattices. They can be combined with predefined models and lattices made in the simple mode. In the following, the usage of the simple mode is shown. For details on using the standard mode, see the online manual for DSQSS/DLA\cite{dsqss}.

\subsubsection{Preparing input files}
The executable "dla", which is the main solver of the DSQSS/DLA, requires the following input files:
(i) parameter file, (ii) lattice file, and (iii) algorithm file.
Here, \verb|dla_pre| is a tool for generating these files from an input file, i.e., "std.toml" in Fig. 3. A user must edit this input file, which contains the following five sections:
\begin{enumerate}
\item{\verb|hamiltonian| }
 In \verb|dla_pre|,  two types of models are predefined. One is the XXZ model:
\begin{align}
\begin{split}\ham &= \sum_{\langle i, j \rangle} -J_z S_i^z S_j^z -\frac{J_{xy}}{2} \left( S_i^+ S_j^- + S_i^- S_j^+ \right)\nonumber\\
&+ D \sum_i \left(S_i^z\right)^2 - h \sum_i S_i^z\end{split}
\end{align}
The other is the Bose-Hubbard model:
\begin{align}
\begin{split}\ham &= \sum_{\langle i, j \rangle} \left[ -t b_i^\dagger \cdot b_j + h.c. + V n_i n_j \right] \nonumber\\
&+ \sum_i \left[ \frac{U}{2} n_i(n_i-1) - \mu n_i \right]\end{split}
\end{align}
In this section, parameters for constructing the above models, such as $J_z$ and $t$, are specified.

\item{\verb|lattice| }
The parameters for the lattice, such as the lattice type \verb|lattice|, the number of dimensions  \verb|dim|, the number of sites along each direction \verb|L|, and the boundary condition \verb|bc|, are specified.
\item{\verb|parameter|}
Parameters for the MC calculation, such as the inverse temperature \verb|beta|, the set of MC sweeps \verb|nset|, the number of MC steps for estimating hyperparameter  \verb|npre|, the number of MC sweeps for thermalization \verb|ntherm|, the number of MC sweeps for measurement \verb|nmcs|, and the seed for the random number generator \verb|seed|, are specified.
\item{\verb|kpoints|}
The interval for measurements in wavenumber space is specified by \verb|ksteps|.
\item{\verb|algorithm|}
The algorithm for calculating the scattering probability for the worm heads is specified by \verb|kernel|.
The following kernels are available:\\
(i) \verb|suwa todo| \\
Rejection minimized algorithm without detailed balance condition (irreversible) proposed by Suwa and
Todo~\cite{Suwa2010}. \\
(ii) \verb|reversible suwa todo| \\
Rejection minimized algorithm with detailed balance condition (reversible) proposed
by Suwa and Todo~\cite{Suwa2013,Todo2013}. \\
(iii) \verb|heat bath| \\
Heat bath method (Gibbs sampler)~\cite{Creutz_1980}.\\
(iv) \verb|metropolis| \\
Metropolis-Hastings algorithm~\cite{Metropolis_1953, Hastings_1970}.
\end{enumerate}
In the following, as the simplest example, the input file for an antiferromagnetic Heisenberg dimer is shown (located in \verb|sample/dla/01_spindimer/std.toml|):
\begin{verbatim}
[hamiltonian]
model = "spin"
M =  1                 
Jz = -1.0              
Jxy = -1.0           
h = 0.0   

[lattice]
lattice = "hypercubic" 
dim = 1                
L = 2                  
bc = false           

[parameter]
beta = 100             
nset = 5              
npre = 10             
ntherm = 10            
nmcs = 100             
seed = 31415          
\end{verbatim}

After preparing the input file, \verb|dla_pre| can be run by typing the following command:
\begin{verbatim}
$ dla_pre std.toml
\end{verbatim}
The following five files are generated:
a parameter file \verb|param.in|, a lattice file \verb|lattice.xml|, an algorithm file \verb|algorithm.xml|, a wavevector file \verb|wv.xml|, and a relative coordinate file \verb|disp.xml|. 
While using the DSQSS/DLA in the simple mode, it is not necessary to understand these intermediate files, since arbitrary lattice systems can be studied by editing these files directly. 
The parameter file \verb|param.in| is a simple text file that specifies the temperature, the conditions for Monte Carlo sampling, such as the number of Monte Carlo steps, and the filenames of the output data.
The \verb|lattice.xml| file defines the lattice structure in XML format. 
The \verb|algorithm.xml| file defines the structure of vertices, the scattering probability matrices, and the tables of vertex states after worm scattering, in XML format.
The \verb|wv.xml| file defines the wavevectors used to calculate several observables: staggered magnetization, dynamical structure factors, and momentum space temperature  Green's functions.
The \verb|disp.xml| file defines the relative coordinates between two sites to calculate real-space temperature Green's functions.

By editing these XML files, DSQSS/DLA can calculate various types of lattices and models, but these XML files are a little complicated to edit.
DSQSS/DLA offers easier files for defining lattice and Hamiltonian
(see the "standard mode" in the Figure \ref{fig:flow_dsqss}).
For example, DSQSS/DLA can deal with the SU($N$) antiferromagnetic Heisenberg model by editing \texttt{hamiltonian.toml}.
For more detailed instructions for editing these intermediate files, we refer the reader to the manual and sample~\cite{dsqss}.

\subsubsection{Performing quantum Monte Carlo calculations}

Once the input files for \verb|dla| are prepared, we can perform a quantum Monte Carlo calculation based on the DLA using the DSQSS/DLA by typing the following command:
\begin{verbatim}
$ dla param.in
\end{verbatim}
Random number parallelization using MPI can be performed by typing the following command:
\begin{verbatim}
$ mpiexec -np 4 dla param.in
\end{verbatim}
Using the above command, the total number of MC samples is multiplied by four (equal to the number of processes), and the statistical error is expected to decrease by half (equal to the inverse square root of the number of processes). 
\subsubsection{Analyzing results}
The results of the calculations are written in the text file \verb|sample.log|. 
This file contains main results such as the sign of the weights \verb|sign|, the energy density (energy per site) \verb|ene|,  the specific heat \verb|spe|, the transverse susceptibility \verb|xmx|, the magnetization \verb|amzu|, and the structure factor \verb|smzu|.
From this file,  the energy can be displayed using the grep command:
\begin{verbatim}
$ grep ene sample.log
R ene = -3.74380000e-01 5.19493985e-03
\end{verbatim}
The two values are the expectation value and the statistical error, respectively. The resulting value is compatible $-3|J|/8=-0.375|J|$ with the exact solution within the statistical error.

\subsection{Usage of the Discrete Space Quantum Systems Solver/parallelized multiple-worm algorithm (DSQSS/PMWA)}

The DSQSS/PMWA can perform QMC calculations for the $S=1/2$ antiferromagnetic XXZ model with nearest-neighbor exchange interactions (or a hard-core boson model with nearest-neighbor repulsive interactions) defined on a one-dimensional chain, a square lattice, and a cubic lattice.
The DSQSS/PMWA implements a parallelized multiple-worm algorithm for large-scale calculations on massively parallel computers.

For QMC simulations by the DSQSS/PMWA, the user needs to prepare the lattice file \verb|lattice.xml| and the parameter file \verb|param.in|.
These two files can be generated by  \verb|pmwa_pre|.
For example, the content of the input file \verb|std.in| for parallelized QMC simulations of the one-dimensional eight-site Heisenberg model at the inverse temperature $\beta = 10.0$ is
\begin{verbatim}
[System]
solver = PMWA
[Hamiltonian]
model_type = spin
Jxy = -1.0
Jz = -1.0
Gamma = 0.1
[Lattice]
lattice_type = square
D = 1
L = 8
Beta = 8.0
[Parameter]
runtype = 0
cb = 1
seed = 31415
nset = 10
nmcs = 10000
npre = 10000
ntherm = 10000
ndecor = 10000
latfile = lattice.xml
outfile = sample.log
nldiv = 2
nbdiv = 1
\end{verbatim}
In the last two lines, the numbers of spatial and temporal divisions are specified for parallel computing.
In this example, the eight-site lattice is divided into two four-site sublattices using the parameter \verb|nldiv| so that a parallelized computation can be performed by two CPUs.
For details of the parameters specified in \verb|std.in|, see the manual\cite{dsqss}.

In order to generate the lattice file \verb|lattice.xml| and the parameter file \verb|param.in|, execute the following command:
\begin{verbatim}
$ pmwa_pre std.in
\end{verbatim}
The DSQSS/PMWA can use the XXZ model \begin{align}
\ham &= - \sum_{\langle i, j \rangle} \left( J_z S_i^z S_j^z + \frac{J_{xy}}{2} ( S_i^+ S_j^- + S_i^- S_j^+ )\right)\nonumber\\
&- h \sum_i S_i^z - \Gamma \sum_i S^x_i
\end{align}
or the hard-core boson model
\begin{align}
\ham &= \sum_{\langle i, j \rangle} \left[ -t b_i^\dagger \cdot b_j + h.c. + V n_i n_j \right] \nonumber\\
&-\mu \sum_i n_i - \Gamma \sum_i (b_i + b_i^\dagger).
\end{align}
The parameters for these models are specified in the input file \verb|std.in|.
Note that the term for the transverse magnetic field (or the source term for the boson particle) is added to the Hamiltonian.
This term is required in the parallelized multiple-worm algorithm.
For calculation of the zero transverse field $(\Gamma = 0)$, we need to extrapolate the numerical results to $\Gamma = 0$.

After preparing the above files, execute \verb|pwma_H| for the XXZ model as
\begin{verbatim}
$ mpiexec -np 2 pmwa_H param.in
\end{verbatim}
or \verb|pwma_B| for the hard-core boson model as 
\begin{verbatim}
$ mpiexec -np 2 pmwa_B param.in
\end{verbatim}
By typing these commands, the parallelized QMC simulation will start using the spatial and temporal domain divisions specified in the lattice file \verb|lattice.xml|.
In the sample input file described above, the results are output to \verb|sample.log| after executing \verb|pmwa_H|.

\section{Applications}
In the DSQSS, several example application samples are provided.
In this section, we introduce three typical samples: magnetic susceptibility of antiferromagnetic spin chains, dynamic spin structure factor, and number density of hard-core bosons on a square lattice.

\subsection{Magnetic susceptibility of antiferromagnetic spin chains}
First, we show the temperature dependence of the magnetic susceptibilities of spin-$1/2$ and $1$ antiferromagnetic spin chains of $32$ sites with a periodic boundary condition ($J = J_z = J_{xy} = -1$).
The results at different temperatures are obtained independently.
In order to perform the calculations automatically, the Python script in \verb|sample/dla/02_spinchain/exec.py| is used.

\begin{figure}[tb!]
  \begin{center}
    \includegraphics[width=1 \columnwidth]{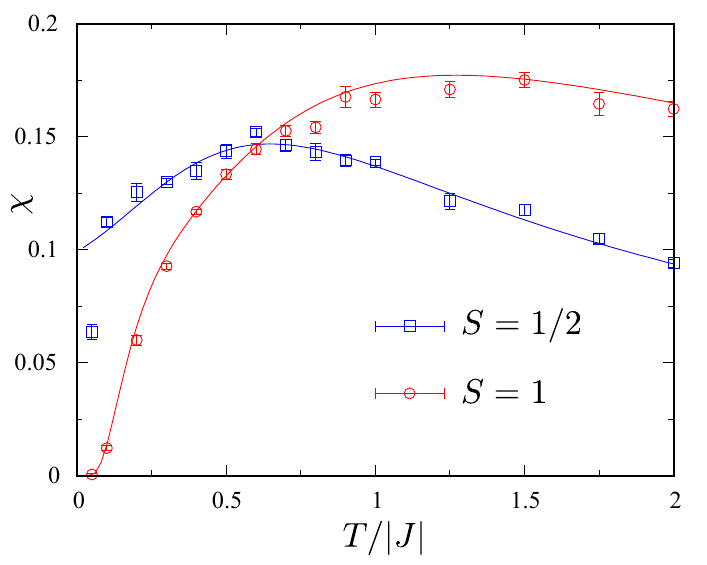}
    \cprotect\caption{Temperature ($T/|J|$) dependence of magnetic susceptibility ($\chi$) for spin-$1/2$ and spin-$1$, respectively. The error bars represent the standard error.
    The solid lines indicate the phenomenological curves~\cite{Bonner1964, Estes1978, Souletie2004}. 
    }
    \label{fig:mag_temp_dep}
  \end{center}
\end{figure}

Figure \ref{fig:mag_temp_dep} shows the temperature ($T/|J|$) dependence of the magnetic susceptibility ($\chi$) for spin-$1/2$ (blue squares) and spin-$1$ (red circles).
The error bars represent the standard error.
The blue and red curves are the phenomenological expressions in the thermodynamic limit for the spin-1/2 chain~\cite{Bonner1964, Estes1978} and the spin-1 chain~\cite{Souletie2004}, respectively.
The magnetic susceptibility for spin-$1/2$ is non-zero at a temperature of absolute zero (the shoulder near T = 0 is due to the finite-size  gap and should disappear in the large system size limit), whereas that for spin-$1$ drops to zero at a certain temperature due to the spin gap~\cite{QuantumMagnetism}.

\subsection{Dynamic spin structure factor}
Next, we consider the dynamic spin structure factor for a one-dimensional spin-$1/2$ Heisenberg chain of $32$ sites with a periodic boundary condition ($J_z = J_{xy} = 1$). In the DSQSS, imaginary-time spin correlation functions can be calculated.
The input file for the DSQSS/DLA for this calculation can be found  in \verb|sample/dla/04_spindynamics|.
The dynamic spin structure factor can be obtained using numerical analytic continuation.
Some source codes are available for numerical analytic continuation~\cite{Maxent, ANN2020, Yoshimi2019}. 
In the DSQSS, a simple code based on the Pad\'e approximation is provided in \verb|sample/dla/04_spindynamics|. 

\begin{figure}[tb!]
  \begin{center}
    \includegraphics[width=1 \columnwidth]{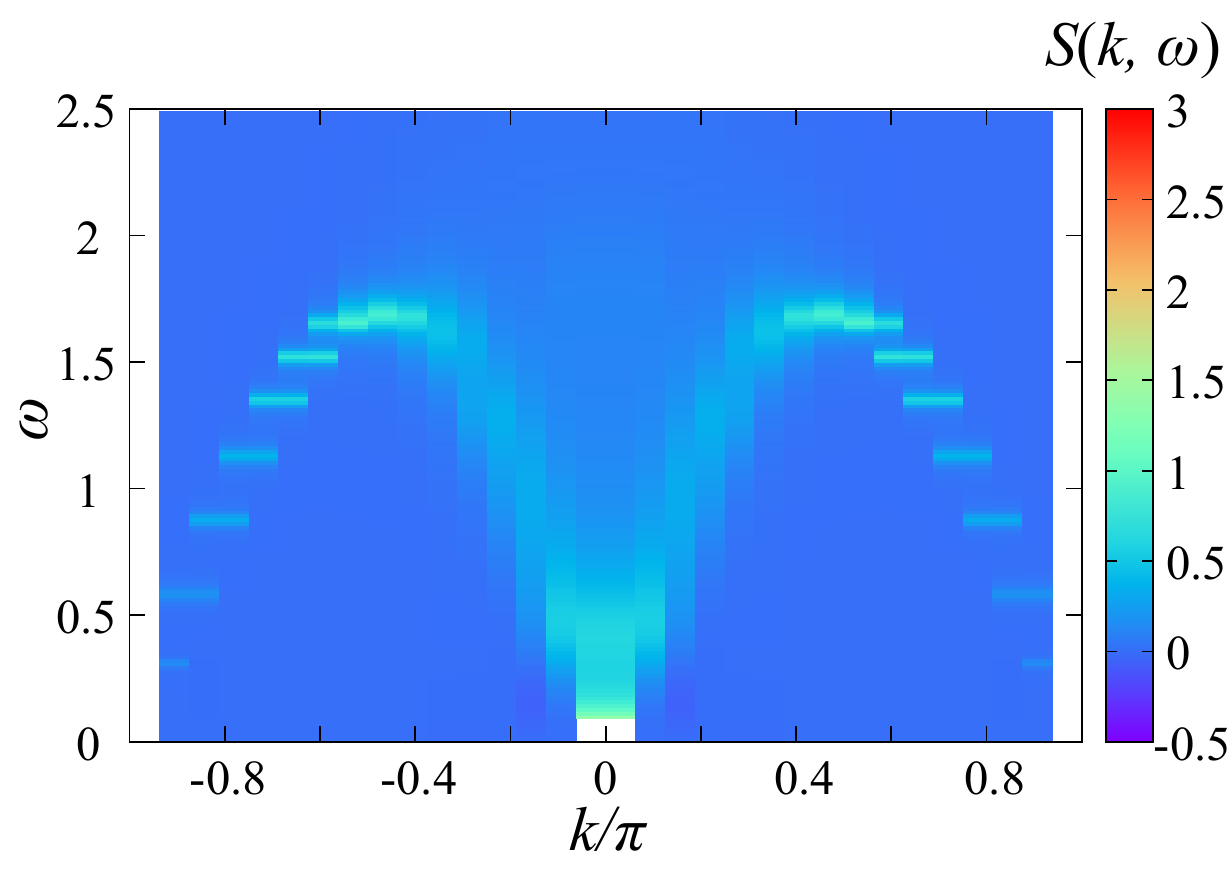}
    \cprotect\caption{Heat map of the spin structure factor $S(k,\omega)$ for the one-dimensional Heisenberg model ($N=32$ and $\beta = 16$). The horizontal axis is the normalized wave vector $k/\pi$, whereas the vertical axis is the real frequency $\omega$. }
    \label{fig:spindynamics}
  \end{center}
\end{figure}

In Fig.~\ref{fig:spindynamics}, we show a heat map of the dynamical spin structure factor $S(k, \omega)$ at the inverse temperature $\beta = 16$. The des Cloizeaux-Pearson mode $E_k =(\pi J/2)\sin(k)$ is seen to appear~\cite{desCloizeaux1962}.
Although dynamic structure factors can also be calculated by the exact diagonalization method using other program packages~\cite{ALPS,Albuquerque2007,Bauer2011,Kawamura2017,HPhi}, these factors can only be used for a small number of sites.
Therefore, QMC codes such as the DSQSS has a remarkable advantage when simulating a large number of spins is necessary, e.g., higher-dimensional cases.
However, note that the dynamic structure factor generally requires highly accurate QMC data using a large degree of sampling because the numerical analytic continuation is an ill-defined problem~\cite{Gubernatis1991,Otsuki2017, Otsuki2020}.
Recent QMC calculation of the spectrum of two-dimensional spin systems can be found in~\cite{Shao2017,su2020stable}.

\subsection{Number density of hard-core bosons on a square lattice}
As an example of a calculation for a boson system,
the chemical potential dependence of the number density $n$ for the hard-core Bose-Hubbard model with repulsive nearest neighbors on an \(8\times8\) square lattice ($t=1$ and $V=3$) is described.
The results are obtained by calculating the number density for different chemical potentials. 
The Python script for performing this calculation automatically is \verb|sample/dla/03_bosesquare/exec.py|.

\begin{figure}[tb!]
  \begin{center}
    \includegraphics[width=1 \columnwidth]{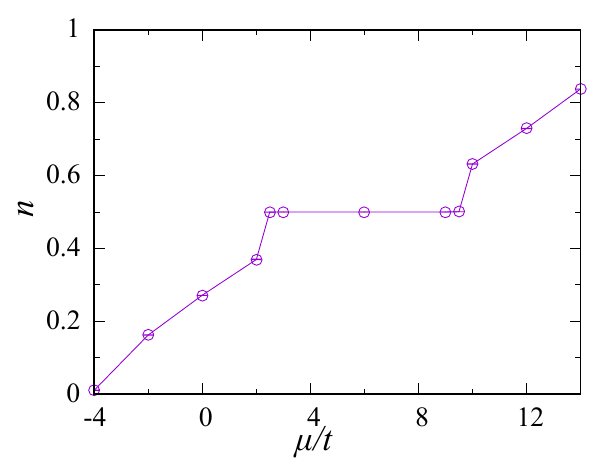}
    \cprotect\caption{Chemical potential ($\mu/t$) dependence of the number density $\mu$ for repulsive hard-core bosons. The error bars represent the standard error, but are almost too small to be seen.}
    \label{fig:chem_boson}
  \end{center}
\end{figure}

Figure~\ref{fig:chem_boson} shows the chemical potential dependence of the number density at an inverse temperature $\beta = 10.0$.
A plateau is seen to be present around \(\mu=6\).
In this region, a checker board solid phase due to repulsive interaction appears~\cite{Batrouni_2000, Hebert_2001, Schmid_2002}.

\section{Summary}
In the present paper, we introduced the Discrete Space Quantum Systems Solver, DSQSS, which performs quantum Monte Carlo simulations based on the directed-loop algorithm.
We described the QMC algorithm used in this program package, as well as how to install and execute this algorithm.
We also presented tutorials on QMC simulations for lattice models with interacting spins and bosons.

The DSQSS is easy-to-use software with a simple user interface, which has several unique features.
The DSQSS/DLA performs QMC calculations of imaginary-time correlation functions, which can be related to dynamical quantities such as the dynamical structure factor.
The DSQSS/PMWA implements an efficient QMC algorithm based on temporal and spatial domain division, which is suitable for massively parallel computing.
These functionalities of the DSQSS will help the user in evaluating physical quantities without complex coding and will promote the user's research activities.

\section*{Acknowledgments}
We would like to thank Yasuyuki Kato, Kota Sakakura, Takafumi Suzuki, Kenji Harada, and Tsuyoshi Okubo for their cooperation in developing the DSQSS.
KY and YM were supported by Building of Consortia for the Development of Human Resources in Science and Technology, MEXT, Japan.
KY was supported by JSPS KAKENHI Grant No. 19K03649, and Building of Consortia for the Development of Human Resources in Science and Technology, MEXT, Japan.
The DSQSS was developed with the support of the "Project for Advancement of Software Usability in Materials Science (PASUMS)" in fiscal year 2018 by the Institute for Solid State Physics, University of Tokyo.
The computations in the present study were performed using the facilities of the Supercomputer Center, Institute for Solid State Physics, University of Tokyo.

\appendix

\section{Measurements of physical quantities in the Discrete Space Quantum Systems Solver (DSQSS)}

\subsection{Diagonal operators}

For diagonal operators such as the total magnetization $\sum_i S_i^z$ and its Fourier transformation with wave vector
$\sum_i S_i^z e^{-i \vec{r}_i\cdot\vec{k}}$, the corresponding estimators are simply given as
\begin{equation}
    A_1(c) = A(\tau = 0) = \frac{1}{N}\Braket{\phi_1 | A |\phi_1}.
\end{equation}
Averaging over imaginary time
\begin{equation}
\begin{split}
    A_2(c) &= \frac{1}{N\beta} \int_0^\beta d\tau \Braket{\phi(\tau)|A|\phi(\tau)} \\
    &=
    \frac{1}{N\beta} \sum_k (\tau_{k+1} - \tau_k) \Braket{\phi_k|A|\phi_k} 
\end{split}
\end{equation}
may improve the statistics.
The DSQSS calculates the diagonal operators using both estimators.
Using the fluctuating-dissipating theorem, the static structure factor and the susceptibility are calculated as: \begin{align}
    S &= N \left( \Braket{A_1^2} - \Braket{A_1}^2 \right), \\
    \chi &= N\beta \left( \Braket{A_2^2} - \Braket{A_2}^2 \right),
\end{align}
respectively.
The DSQSS calculates the imaginary-time correlation functions for the diagonal operator as
\begin{equation}
C(\vec{k}, \tau) = \frac{1}{\beta}\int_0^\beta d\tau' \Braket{A(\vec{k}, \tau+\tau')A(-\vec{k}, \tau')},
\end{equation}
for the discretized imaginary times.
In this calculation, $Ae^{\sum_j -i \vec{r}_j \cdot \vec{k}}$ is measured for each time slice and each $\vec{k}$-point.

\subsection{Energy and specific heat}

The total energy (the expectation value for the Hamiltonian) can be calculated as the logarithmic derivative of the partition function with respect to the inverse temperature:
\begin{equation}
    E = \Braket{\ham}  = -\partial_\beta \log Z
    =
    -\frac{\sum_c \partial_\beta W(c)}{\sum_c W(c)}.
\end{equation}
Substituting Eq. (\ref{eq:weight}) and changing variables $\tau = \beta \tau'$ yields the following estimator:
\begin{equation}
    E(c) = \frac{1}{\beta}\left( \int_0^\beta d\tau \ham_0(\tau)\right) - \frac{\braket{n}}{\beta},
\end{equation}
where $n$ is the order of perturbations.
The estimator for the specific heat per site is given as
\begin{equation}
\begin{split}
    C &= \frac{1}{N}\partial_T E = \frac{\beta^2}{N}\partial_\beta E \\
    &=
     \frac{1}{N}\Braket{\left( \int_0^\beta d\tau \ham_0(\tau) - n\right)^2} \\
     &\quad - \frac{1}{N}\Braket{ \int_0^\beta d\tau \ham_0(\tau) - n}^2
     - \frac{\braket{n}}{N} ,
    \end{split}
\end{equation}
where $N$ is the number of sites.

\subsection{Helicity modulus}

The helicity modulus~\cite{Fisher1973}, which corresponds to the spin stiffness in spin models and to the superfluid density in bosonic models, is measured as the susceptibility with respect to the twist of spins $\theta$ along the $\mu = x,y,z$ axis,
\begin{equation}
    \rho_s = -\frac{1}{\beta L_\mu^2}\partial_\theta^2 \log Z,
\end{equation}
where $L_\mu$ is the length of the system along the $\mu$ axis.
The DSQSS calculates the helicity using winding numbers of world-lines~\cite{Pollock_1987, Ceperley_1989} as
\begin{equation}
    \rho_s = \frac{1}{\beta d V} \sum_\mu^{x,y,z}L_\mu\Braket{W_\mu^2},
\end{equation}
where $d$ is the spatial dimension, and $V = L_x L_y L_z$.

\subsection{Off-diagonal operators}

The DSQSS calculates the temperature Green's functions as
\begin{align}
G(\vec{r}, \tau) &= \frac{1}{N\beta} \sum_{\vec{r}'} \int_0^\beta d\tau'
\nonumber \\
& \hspace{-3mm} \times
\Braket{T_\tau Q(\vec{r}+\vec{r}', \tau+\tau') Q^\dagger(\vec{r}', \tau')}
\\
G(\vec{k}, \tau) &= 
\frac{1}{\beta} \int_0^\beta d\tau' \nonumber \\
& \times \Braket{T_\tau Q^\dagger(\vec{k}, \tau+\tau') Q(-\vec{k}, \tau')},
\end{align}
where $Q$ is an off-diagonal site operator, such as $S^x_i$.
Unlike other observables,
they also need configurations with worm heads.
The correlation $\braket{Q(X)Q(Y)}$, where $X$ and $Y$ are coordinates in $d+1$ dimensional spacetime, is proportional to the ratio of the partition function of the original system to that of the system, where two worm heads exist on $X$ and $Y$:
\begin{equation}
\begin{split}
&    (\eta d\tau)^2 \Braket{Q(X)Q(Y)} \\
    &= \frac{\mathrm{Tr}\mathrm{T}_\tau e^{-\beta \ham} Q(X)Q(Y)}{\mathrm{Tr}e^{-\beta \ham}} \\
    &=
    \sum_{c:X,Y}W(c) \Big/ \sum_{c: \text{No heads}}W(c).
\end{split}
\end{equation}
This ratio can be evaluated as that between the number of configurations appearing in an MC cycle.
Therefore, the temperature Green's function can be evaluated as follows:
\begin{equation}
    G(\vec{r}, \tau)
    =
    \frac{1}{N\beta\eta^2}
    \Braket{n_\text{worm}(\vec{r}, \tau)},
\end{equation}
where $n_\text{worm}(\vec{r}, \tau)$ is the number of times the distance between two heads is $(\vec{r}, \tau)$ in a MC cycle.
In addition, $G(\vec{k}, \tau)$ can be estimated in the similar manner.

\bibliographystyle{elsarticle-num}
\bibliography{reference}

\end{document}